\documentclass[twocolumn,showpacs,preprintnumbers,amsmath,amssymb]{revtex4}
\usepackage{graphicx}
\usepackage{dcolumn}
\usepackage{bm}
\usepackage{amsmath}

\newcommand{\bi}{\begin{itemize}}
\newcommand{\ei}{\end{itemize}}
\newcommand{\bt}{\begin{theorem}}
\newcommand{\et}{\end{theorem}}
\newcommand{\bp}{\begin{proof}}
\newcommand{\ep}{\end{proof}}
\newcommand{\be}{\begin{equation}}
\newcommand{\ee}{\end{equation}}
\newcommand{\ben}{\begin{enumerate}}
\newcommand{\een}{\end{enumerate}}

\newcommand{\C}{\mathbb C}

\newcommand{\R}{\mathbb R}

\newcommand{\Rscr}{\mathcal R}

\newcommand{\hf}{\frac{1}{2}}

\newcommand{\e}{\varepsilon}

\newcommand{\m}{\mu}

\renewcommand{\O}{\Omega}
\renewcommand{\o}{\omega}

\renewcommand{\t}{\tau}
\renewcommand{\b}{\beta}

\newcommand{\z}{\zeta}
\newcommand{\x}{\xi}
\newcommand{\D}{\Delta}
\renewcommand{\S}{\Sigma}
\renewcommand{\a}{\alpha}

\newcommand{\g}{\gamma}

\renewcommand{\and}{\text{~~ and ~~}}
\renewcommand{\part}{\partial}
\newcommand{\ra}{\rightarrow}
\newcommand{\sech}{{\rm sech~}}

\newcommand{\ipt}{\frac{i\pi}{2}}

\renewcommand{\D}{\Delta}

\begin{document}

\title{Temporal evolution of attractive  Bose-Einstein condensate in a quasi 1D cigar-shape trap
modeled through the semiclassical limit of the focusing Nonlinear Schr\"{o}dinger Equation}
\author{Alexander Tovbis}
\affiliation{Department of Mathematics, University of Central Florida, Orlando, Florida 32816}
%\author{Viatcheslav Kokoouline}
%\affiliation{Department of Physics, University of Central Florida, Orlando, Florida 32816}

%\ead{douguet@physics.ucf.edu}

\begin{abstract} 
One-dimensional (1D) Nonlinear Schr\"{o}dinger Equaation (NLS) provides a good approximation to 
 attractive Bose-Einshtein condensate (BEC) in a quasi 1D
 cigar-shaped optical trap in certain regimes. 1D NLS is an integrable equation
that can be solved through the inverse scattering method. Our observation is that in many cases
the parameters of the BEC correspond to the semiclassical (zero dispersion) limit of the 
focusing NLS. Hence, recent results about the {\it strong asymptotics}  of the semiclassical limit solutions 
can be used  to describe some interesting phenomena  of the attractive 1D BEC.
In general, the semiclassical limit of the focusing NLS exibits very strong 
modulation instability. However, in the case of an analytical initial data, the NLS evolution  does
displays some  ordered structure, that can describe, for example, the bright soliton phenomenon.
We discuss some general features of the semiclassical NLS evolution and propose some new observables
to the attractive 1D BEC.

\end{abstract}

\pacs{}

\maketitle

{\bf 1D BEC as a semiclassical limit of the focusing NLS.}
It is generally accepted that the temporal evolution of the BEC   is governed by the Gross-Pitaevskii
(GP) equation   for the condensate wave function $\Psi(\bf{r},t)$, given by
\begin{equation}
\label{GPEfull}
i\hbar \frac{\partial}{\partial t}\Psi=\left[-\frac{\hbar^2}{2m}\nabla^2 +V_{ext}({\bf r})+g\vert\Psi\vert^2\right]\Psi\,,
\end{equation}
where:  $m$ is the single atom mass; $V_{ext}$ is an external trapping potential that
we consider to be
%can be taken as 
a harmonic oscillator potential, 
and  $g$ is the coupling constant determined by the scattering length.
%: $g=4\pi\hbar^2 a_s/m$, $a_s$ being the s-wave scattering length. 
Negative values of $g$ correspond to the attractive BEC.
If the characteristic energies of the radial excitations are much greater then the energy of the nonlinear term
(see estimate \eqref{1Dcond} below),
the 3D GP equation \eqref{GPEfull} can be approximated by a 1D GP in the longitudial (axial)
direction (\cite{Abdul})
\begin{equation}
\label{GPE1D}
\left[i\hbar \frac{\partial}{\partial t}+\frac{\hbar^2}{2m}\frac{\partial ^2 }{\partial \x^2}-V_{ext}({\x})-\frac{g}{2\pi l^2_\perp}\vert\psi(\x,t)\vert^2\right]\psi(\x,t)=0,
\end{equation}
where $\x$ is the axial variable and 
\be\label{extpot}
V_{ext}({\x})=\hf m \o_\x \x^2
\ee
 is the axial part of the harmonic trapping potential. 
Here $g=\frac{4\pi \hbar^2 a_s}{m}$, $l^2_\perp=\frac{\hbar}{m\o_\perp}$, $\o_\perp$ and  $\o_\x$ are  the 
trap frequencies in the radial and the axial  directions respectively. Taking as an example  the bright solitons
experiment with Lithium  ($^7$Li, Sreckker et al., \cite{Streck}), we have 
$m\approx 10^{-26}$kg,
$a_s\approx -3a_0\approx -16\cdot 10^{-11}$m (here $a_0$ denotes Bohr radius) and   
$\o_\perp = 2\pi\cdot 640$Hz$\approx 4\cdot 10^3$H;  in the first approximation,
we put  $\o_\x=0$.
The axial wave function $\psi(\x,t)$ is assumed to have a shape of Gaussian and is normalized by
\be\label{normatom}
\int_\R |\psi(\x,0)|^2 d\x =N~,
\ee
where $N\approx 3\cdot 10^5$ is the total number of the atoms in the trap. 
%Following the setup of the bright soliton experiment of \cite{Streck} (see also \cite{Abdul}), we assume $N \approx 3\cdot 10^5$.

Considering 
%\be\label{unifdensity}
$|\Psi|^2\sim \frac{N}{l^2_\perp s_\parallel}$,
the
1D GPE \eqref{GPE1D} is applicable (\cite{Abdul}) under the condition
\be\label{1Dcond}
\frac{N|a_s|}{s_\parallel}\ll 1,
\ee
where $s_\parallel$ is the order of magnitude of the size of the condensate in the axial direction.
Assuming $s_\parallel\approx 3\cdot 10^{-4}$m, (\cite{Streck}), the left hand side of \eqref{1Dcond} becomes
$0.16$, which may give some justification for the use of \eqref{GPE1D}.

Equation \eqref{GPE1D} with zero external potential is a 1D NLS, which can be integrated through
the inverse scattering technique. 
We start our discussion by showing that 
%a solution $\psi(r,t)$ to 
the GP equation \eqref{GPE1D} 
with zero external potential that describes the attractive BEC in a cigar-shaped trap can be rescaled
to 
%We discuss the applicability of the assumption and how it can be avoided later. 
%In the first step, we reduce \eqref{GPE1D}  to
\begin{equation}
\label{FNLS}
i\varepsilon q_\t +\left(\frac{\varepsilon^2}{2}\right)q_{xx}+\vert q\vert^2 q=0~,
\end{equation}
 where $\e$ is a small positive parameter,  $x$, $\t$ are scaled space - time variables and
the initial data $|q(x,0)|$ has a shape of Gaussian with a typical lenght of order 1 is normalized by
\be\label{normNLS}
\int_\R |q(x,0)|^2 dx =1~.
\ee
Equation \eqref{FNLS} is the standard form of the focusing NLS in the semiclassical (zero dispersion) limit.

Substitution of expressions for $g,l_\perp$ into \eqref{GPE1D} yields
\be\label{GPE1Dred}
\left[ i \frac{\partial}{\partial t}+\frac{\hbar}{2m}\frac{\partial ^2 }{\partial \x^2}+2|a_s|\o_\perp\vert\psi(\x,t)\vert^2\right]\psi(\x,t)=0.
\ee
Equation \eqref{FNLS} can be obtained from \eqref{GPE1Dred}
through the change of variables
\be\label{chvar}
\psi(\x,t)=\b q(x,\t),~~~~~\x=\D x, ~~~~~~t=k\t~, 
\ee 
where the coefficients $\b,\D$ and $k$ are to be determined.
The comparison of the norming conditions \eqref{normatom} and \eqref{normNLS}
yields $\b^2=\frac{N}{\D}$.
Substituting \eqref{chvar} into \eqref{GPE1Dred} one gets
\begin{equation}
\label{NLSder1}
i\frac{\D}{2|a_s|\o_\perp Nk}q_\t +\hf \left(\frac{\hbar}{2|a_s|\o_\perp Nm\D}\right)q_{xx}+\vert q\vert^2 q=0~.
\end{equation}
Comparison of equations \eqref{NLSder1} and \eqref{FNLS} yields
\be\label{const1}
\e=\frac{\D}{2|a_s|\o_\perp Nk}~~~~~{\rm and}~~~~~~\frac{\D^2}{2|a_s|\o_\perp Nk^2}=\frac{\hbar}{m\D}~,
\ee
%Immediate consequences of \eqref{const1} are
so that
\be\label{const2}
k=\D\sqrt{\frac{m\D}{2|a_s|\o_\perp N\hbar}}~~~~~{\rm and}
~~~~~~\e=\frac{\sqrt{\hbar}}{\sqrt{2|a_s|\o_\perp Nm\D }}~.
\ee
Comparison of the typical size of Gaussian distributions for $|\phi(\x,0)|$ and $|q(x,0)|$
gives
\be\label{Dval}
\D\sim 10^{-4}~.
\ee
Using numerical value $\hbar\approx 10^{-34}\frac{{\rm m}^2{\rm kg}}{\rm s}$,
%numerical values of
%\be\label{mhbar}
%m\approx 10^{-26}{\rm kg~~~~~and~~~~~~}
we calculate
\be\label{orders}
\e\approx 1.6\cdot10^{-2}~~~~~{\rm and}~~~~~~k\approx 5\cdot10^{-3}~.
\ee 
Equation \eqref{orders} shows that the time evolution of attractive BEC, governed  by equation \eqref{GPE1Dred},
can be described by the semiclassical limit of the focusing NLS \eqref{FNLS} with normalization
\eqref{normNLS}. (Normalization of $\int|q|^2dx=n$ instead of \eqref{normNLS}, where $n=O(1)$, leads
to the replacement of $N$ by $N/n$ in equations \eqref{NLSder1}-\eqref{const2}. Thus, in this case
$\e,k$ has to be multiplied by $\sqrt{n}$.)  

{\bf Semiclassical limit solutions to the focusing NLS \eqref{FNLS}.}
The focusing Nonlinear Schr\"odinger  equation \eqref{FNLS},
where $x\in\R$ and $\t\ge 0$ are space-time variables,
is  a basic model for self-focusing and self-modulation;
it describes the evolution of the envelope of modulated wave in 
general nonlinear systems. It is also one of the most celebrated 
nonlinear integrable equations that was first integrated by Zakharov and Shabat \cite{ZS},
who produced a Lax pair for it and used the inverse scattering
procedure to describe general decaying solutions ($\lim_{|x|\to 0}q(x,0)=0$) in terms of radiation  and solitons. 

\begin{figure}
\includegraphics[width = 0.49\textwidth]{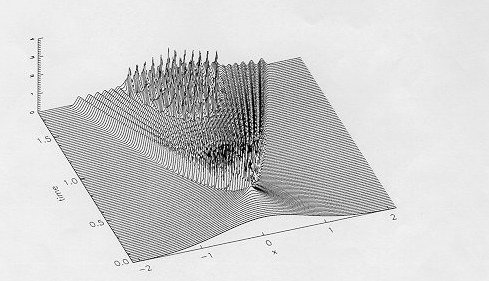}
\caption{Absolute value $|q(x,\t,\varepsilon)|$ of a solution 
$q(x,\t,\varepsilon)$ to the focusing NLS (\ref{FNLS})
versus $x,\t$ coordinates from \cite{CMM}. Here $A(x)=e^{-{x^2}}$, $S'(x)=-\tanh x$ and $\e=0.02$.}
\label{Cai}\end{figure}

In the  semiclassical
limit ($\varepsilon \to 0$) the focusing  NLS \eqref{FNLS}  exibits   
{\it modulationally unstable} behavior (see Fig. \ref{Cai}), as was first shown  in \cite{FL}. 
This is in drastic contrast to the case of the defocusing
NLS equation (\cite{CMM}, \cite{KKU}) 
in which the semiclassical theory shows regions of modulated
periodic or quasiperiodic oscillation. These two very different types
of behavior can be explained through  modulation equations, which are elliptic 
in the focusing and hyperbolic in the defocusing cases. The corresponding initial value problems 
are, therefore, ill-posed and   well-posed respectively. 
As a result, a plane wave
with amplitude modulated by $A(x)$ and phase modulated by $S(x)$, taken as an initial data  
\be \label{IDe}
q(x,0,\varepsilon)=A(x)e^{iS(x)/\varepsilon}
\ee
for the focusing NLS \eqref{FNLS},
is expected to  break immediately into  disordered oscillations 
of both the amplitude and the phase. 
However, in the case of an {\em analytic} initial data, the NLS evolution  
displays
some  orderly structure 
instead of the disorder suggested by the modulational
instability, see   \cite{CMM},  \cite{MK} and 
\cite{CT}. Throughout this work, we will use the abbreviation NLS to mean ``focusing Nonlinear Schr\"odinger equation".

Fig. \ref{Cai} from \cite{CMM} depicts the time evolution of a typical Gaussian-shaped symmetrical
analytic initial data \eqref{IDe}.  It clearly identifies regions {where}  different types of  behavior 
of {the} solution $q(x,\t,\varepsilon)$ appear. These regions are separated by some  
independent of $\varepsilon$ curves in the $x,\t$ plane that are called {\rm breaking curves} or {\rm nonlinear caustics}. 
Within each region, the {\em strong asymptotics}  of $q(x,\t,\varepsilon)$ can be expressed in terms of Riemann
Theta-functions (see, for example, \cite{TVZ1}). In this context, regions of different asymptotic behavior of 
$q(x,\t,\varepsilon )$ corresponds to the different genera of the Schwarz symmetrical hyperelliptic Riemann surface 
$\Rscr(x,\t)$, whose Theta-functions enter in the asymptotic description. In the genus zero case, 
$\Rscr(x,\t)$ has only two branchpoints $\a=\a(x,\t)$ and $\bar\a$, so that the 
Riemann Theta-functions expression for  $q(x,\t,\varepsilon)$ is replaced  by
\be\label{qg0}
q(x,\t,\varepsilon)=A(x,\t)e^{iS(x,\t)/\varepsilon},
\ee
where
\be\label{alph}
\a(x,\t)=-\hf S_x(x,\t)+i A(x,\t).
\ee
 The  genus zero region is
the first region adjacent to the axis $\t=0$, where (genus zero)  solution \eqref{qg0} has the form
 of a high frequency ($O(1/\e)$) modulated wave  with  the slowly varying 
amplitude $A(x,t)$ and 
phase $S(x,t)$.  (We remark that $A(x,0)=A(x)$, $S(x,0)=S(x)$.)

In some very few special cases, for example, when $A(x)=\sech x$ and $S'(x)=-\frac{\mu}{2}\tanh x$, $\m\ge 0$, the scattering 
data for \eqref{FNLS} can be calculated explicitly. Then the modulated amplitude and phase of \eqref{qg0}, \eqref{alph}
can be obtained from the system of trancendental equations for $\a(x,\t)=a(x,\t)+ib(x,\t)$:
\begin{equation}
\label{transc}
\begin{cases}
&\sqrt{(a-T)^2+b^2}+\sqrt{(a+T)^2+b^2}=\m+4\t b^2 \\
&\left[a-T+\sqrt{(a-T)^2+b^2}\right]\left[a+T+\sqrt{(a+T)^2+b^2}\right]\\
&=b^2 e^{2(x+4\t a)}~,
\end{cases}
\end{equation}
where $T=\sqrt{\frac{\m^2}{4}-1}$. In the particular case $\m=2$ (the borderline value of $\m$ between
the pure radiational case  $\m>2$ and radiation with solitons case $\m<2$), introducing the implicit
time $u=u(x,\t)$ at each point  $x\in\R$ by $\t=\frac{(u-x)[\sinh 2u-(u-x)]}{8\sinh^2 u}$, one can obtain
an explicit solution (\cite{TVZ1})
\be\label{ab2}
a=\frac{2\sinh^2 u}{\sinh 2u-(u-x)}~,~~~~~b=\frac{2\sinh u}{\sinh 2u-(u-x)}~
\ee
for $A(x,\t)=b$ and $S'(x,\t)=-2a$. Similar expressions are available for the case $\m=0$.

Notice that the amplitude $A(x,t)$ 
of the solution on Fig. \ref{Cai} at first contracts (accumulates) towards the point of maximum ($x=0$) 
of $|q(x,0)|$ and then suddenly bursts into rapid (order $1/\e$)  and violent oscillations in {\em  amplitude}
(transition to genus two regime).
This is the typical behavior (\cite{TVZ3}) for an analytic one-hump initial data (provided that $S'(x)$ does not 
decrease too fast)
The very first point of this transition, which is the tip-point of the first breaking  curve 
(see  Fig. \ref{Cai}), is called a point of  {\it gradient catastrophe}, or {\it elliptic umbilical
singularity} (\cite{DGC}).  At the  point $(x_0,\t_0)$ of gradient catastrophe 
the  semiclassical solution  
\eqref{qg0} of \eqref{FNLS} losses its smoothness (\cite{TV3}), i.e., 
 $\a_x(x_0,\t_0)=\infty$  (either $A_x(x,\t_0)$ or $S_{xx}(x,\t_0)$
or both become infinite). 

\begin{figure}
\includegraphics[width = 0.49\textwidth]{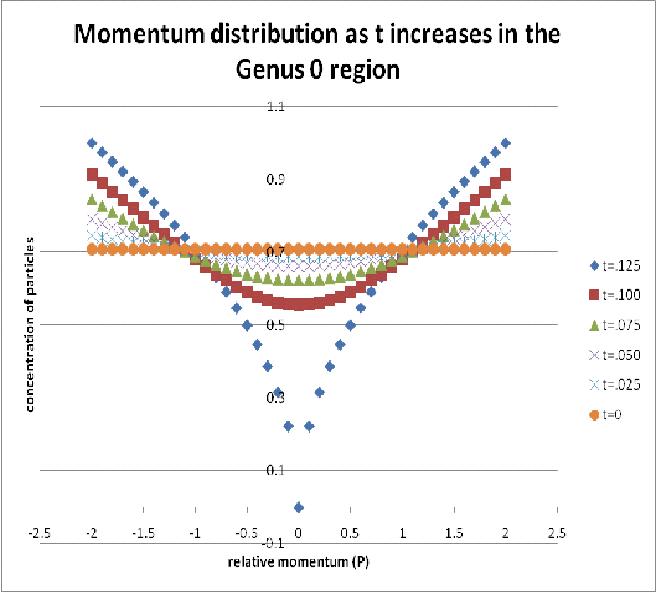}
\caption{Evolution of the Fourier transform of the initial data \eqref{IDe}  with
$A(x)=\sech x$ and $S'(x)=-2\tanh x$ in the limit $\e\ra 0$ from $\t=0$ to the time of gradient catastrophe $\t=0.125$.}
\label{Benfig}\end{figure}

The Theta-function expression for higher genus solutions is somewhat cumbersome for this paper (see,
for example, \cite{TVZ1}).  This expression gives an $O(\e)$ approximation for the solution 
of \eqref{FNLS}, \eqref{IDe} in the corresponding region. 
For a fixed time snapshot in the genus two region (Fig. \ref{Cai},
 around $\t=1$), the graph of $|q(x,\t,\e)|$ can be identified with  bright
solitons, that were  experimentally observed, for example, in \cite{Streck}.
If the BEC with a Gaussian shaped initial data is governed by 1D NLS \eqref{GPE1Dred}, the region filled with solitons 
is  spreading off in the axial direction. 
%(here we do not assume any axial trapping potential). 
According to Fig. \ref{Cai},
the onset of soliton-like (genus two) behavior in the semiclassical limit happens at $\t=\hf$ or
$t=k\t\approx 2.5\cdot 10^{-3}$s, which does not contradict observations of \cite{Streck}, were bright solitons were
first observed at $t=5$ms. Since  the ``effective'' axial  size of the condensate   $s_\parallel$ shrinks considerably
near the time of gradient catastrophe (see Fig. \ref{Cai}), 
 condition \eqref{1Dcond} may be violated during this period.
%
%does not hold near the time of gradient catastrophe, because the ``effective'' axial  size of the condensate   $s_\parallel$ at this time decerases considerably due to the 
%accumulation of the condensate near $x=0$. 
%
This is consistent with the fact that the total  number of atoms
observed in the soliton regime  in the experiment of \cite{Streck} is less than 20\% of the  number of atoms $N\approx 3\cdot10^5$
at the beginning of the experiment.
(The NLS evolution preserves the $L^2$ norm of the solution, i.e., evolution governed by equation \eqref{GPE1Dred}
would  preserve the total number of atoms $N$.)
Moreover, the evolution of the Fourier transform of $q(x,t,\e)$, which can be calculated
explicitly form \eqref{qg0} through the stationary phase method, shows that the portion of atoms 
in the condensate   with high axial momentum significantly increases (see Fig. \ref{Benfig}) as the point of gradient
catastrophe is approached.

Note that  breaking curves (boundaries between the regions of different genera) depend on both the amplitude and the phase  of the initial condition $q(x,0,\e)$, 
but do not depend on $\e$.
Within a region of genus $2n$, $n>0$, the semiclassical limit solution 
can be viewed  as  modulated $2n$-phase nonlinear wave, 
with $2n+1$ complex ($4n+2$ real) wave parameters, which slowly vary (in $x,\t$) 
in the region. (Fig. \ref{Cai} depicts consequitive regions with $n=0,2,4$).
Complex wave parameters can be interpreted as a set of
branchpoints of the Schwarz-symmetrical hyperelliptic surface $\Rscr(x,\t)$, whose evolution in the $x,\t$
plane is defined by modulation (Whitham) equations. (Here we want to mention that regions of genus four
(see Fig. \ref{Cai} after $\t=1.5$)
and of higher genera are associated with the initial data \eqref{IDe} that support solitons, see \cite{TVZ1};
in the semiclassical limit solutions the number of solitons has order $O(1/\e)$.)

{\bf Calculation of semiclassical solutions to \eqref{FNLS}.}
Equation \eqref{FNLS}, as an integrable NLS, can be solved by inverse scattering technique.
However, the semiclassical limit solutions require the {\it semiclassical limit} of the scattering
transform.  Let $z$ be a point on the curve $\S$ in the upper halfplane,
that is defined parametrically by the analytic initial data \eqref{IDe}  as $\a(x)=-\hf S'(x) +iA(x)$, $x\in\R$. 
(Here $A(x)$ and $S'(x)$ have sufficient decay to zero or to some finite
values $\pm \m$ at $\pm\infty$ respectively.)
Assuming for simplicity that $\a(x)$ is invertible, the semiclassical scattering data limit $f_0(z)$, $z\in\S$,
is defined (\cite{TV3}) through  
a generalized  Abel integral transform  as
\begin{align}\label{xtofpart}
 f_0(z)=
\int^{\m_+}_{z}&\left[z-\m_+ + \sqrt{(z-u)(z-\bar u)} \right]x'(u)du\cr
 &+(z-\m_+) x(z),
\end{align}
where $x(\a)$ is inverse to $\a(x)$ and the integral is taken along $\S$. The analytic extension of $f_0(z)$ 
from $\S$ to $\R$ (which can have logarithmic branchcuts) has a meaning of the leading order term
of  $\hf i\e\ln r_0(z,\e)$ as $\e\ra 0$, where  $r_0(z,\e)$, $z\in\R$, is the reflection coefficient of \eqref{IDe}. 
Once $f_0(z)$ is known, the complex wave parameters are defined through the modulation equations. In particular,
in the genus zero region,  the modulation equation for $\a(x,\t)$ is given by
a system of two real equations (\cite{TVZ1})
\begin{equation}
\label{modeqg0}
%\begin{cases}
\int_\gamma \frac{f'(\zeta)}{R_+ (\zeta)} d\zeta =0,~~~~~~~
\int_\gamma \frac{\z f'(\zeta)}{R_+ (\zeta)} d\zeta =0,
%\end{cases}~,
\end{equation}
where $f(z)=f(z;x,t)=f_0(z)-xz-2tz^2$ and  $R(z)=\sqrt{(z-\a)(z-\bar \a)}$.
It defines $q(x,\t,\e)$ through \eqref{qg0}-\eqref{alph}. (Here  $f_0(z)$ is Schwarz symmetrically 
extended into the lower halfplane; typically, $\Im f_0(z)$ has a jump along $\R$.)

Define function $h(z)=h(z;x,\t)$ as
\be\label{h}
h(z)=\frac{1}{i\pi}\int_{\g_m}{\frac{f(\z)}{(\z-z)R(\z)_+}}d\z-f(z)~,
\ee 
where $\g_m$ is  a Shwarz-symmetrical contour connecting $\bar\a$ and $\a$, and such that $\g_m\cup\R=\m_+$.
Because of the analyticity of $f(z)$, a particular shape of $\g_m$ is not important.
However, it is possible to fix $\g_m$ by the condition $\Im h(z)=0$ on $\g_m$.
According to the Deift-Zhou nonlinear steepest descent method, the genus zero anzatz \eqref{qg0}
approximates the actual solution of the NLS \eqref{FNLS}  with the reflection coefficient
$r_0(z,\e)=e^{-\frac{2i}{\e}f_0(z)}$ if  (\cite{TVZ1})
\begin{align}\label{ineq}
\Im h(z;x,\t)&<0 ~{\rm on~ both ~sides~  of~ } \g^+_{m};\cr
\Im h(z;x,\t)&>0 ~{\rm  on~ }\g^+_{c},
\end{align}
where $\g^+_c$ is a contour in the upper halfplane $\C^+$ connecting $\a$ and $\m_-$ and
$\g^+_m=\g_m\cup \C^+$. We have a freedom to deform the contour  $\g^+_c$ so that the inequalities \eqref{ineq} are satisfied
along it.
The first breaking curve consists of points $(x,\t)$ where at least one of the inequalities \eqref{ineq}
turns into equality at some $z_0$. Thus, equation for the first breaking curve can be written as  a system of three
real equations for $z_0\in\C$ and $(x,\t)\in\R^2$
\be\label{br1}
\Im h(z_0;x,\t)=0~~~~~~~~~~{\rm and}~~~~~~~~~~~~h_z(z_0;x,\t)=0~.
\ee

For the initial data \eqref{IDe} with when $A(x)=\sech x$ and $S'(x)=-\tanh x$, the  expression 
\begin{align}\label{hexpl}
&h(z)=
z\ln\frac{\sqrt{a^2+b^2}R(z)-a(z-a)+b^2}{z}-2\t(z-a) R(z)\cr
&+(1-z)\left[\ln b-\ipt\right]-\ln[R(z)-(z-a)]\cr
\end{align}
was found in \cite{TVZ1}.
Modulation equations, as well as expressions for $h(z;x,\t)$ for higher genus regions, can be written in the explicit determinantal form
(see \cite{TV1}, \cite{TV2}), however, since these expressions are somewhat involved, they will not be given in this paper.
%We want to underscore the fact these quantities  can be effectively calculated, so that slowly varying wave paarmeters,
%as well as breaking curves, can be calculated in the cases of higher genus.

{\bf Suggestions and conclusions.}  Semiclassical limit of the focusing 1D NLS \eqref{FNLS} provides
a new, mathematically rigorous tool to study modulationally unstable evolution
of the attractive 1D BEC. Evolution of the BEC with a one-hump initial data that is governed
by the NLS \eqref{GPE1Dred} is expected to show  two or more qualitatively different regimes  
%or  ``phase transitions''
(regions of different genera) within  $O(k)$ (see eq. \eqref{const2})  time interval.
Bright soliton experiment  of \cite{Streck} is an example of  a typical  higher genera region  behavior (apparently, for the genus 2 region),
attained within the time period of $5$ms, where $k\approx 4$ms. 
In general, any macroscopic characteristic of the evolving condensate can be suggested as
an observable. That include (see Fig. \ref{Cai}): 
\bi
\item space-time location of the breaking curves (exact location of a breaking curve is given by \eqref{br1});
\item slowly modulated amplitude $A(x,\t)$ in the genus zero (the exact value of $A(x,t)$ is given through the modulation equation \eqref{modeqg0});
\item the upper and lower envelopes of the high-frequency amplitude oscillations in the genus two region 
(the envelopes are defined through the Riemann-theta functions). 
\ei

\medskip  
Calculation of the observables, mentioned above, is based on  the semiclassical limit of the scattering data $f_0(z)$,
which, in its turn, can be obtained from $q(x,0,\e)$ through  \eqref{xtofpart}, i.e., through the initial amplitude $A(x)$
and the phase $S(x)$. However, accurate measurment of the initial phase is often a difficult task. We can turn the question around and ask
%which leads to the question of 
whether the phase $S(x)$ can be somehow reconstructed from $A(x)$ and some observables. Continuation of this line of argument
leads to the question of designing some NLS-data, initial or scattering, whose evolution will have certain desired properties and/or
fit within some required parameters. Generally speaking, formulae \eqref{xtofpart}-\eqref{br1} are valid for a large class
of analytic initial data, including, for example,  cases of the  multi-hump initial data $A(x)$, and  experimental 
data about evolution of the BEC  with multi-hump  initial density might be  interesting.

\smallskip  
1D NLS approximation of the evolving attarctive BEC in a cigar-shaped trap may be valid during certain intervals
of the total   period of observation and not valid during the others. The  atoms lost in the experiment of \cite{Streck}
(probably, near the point of the gradient catastrophe), seem to indicate that this is an example of such situation. 
Perhaps, some other model  is needed to trace the evolution of the BEC through
the point of the gradient catastrophe into the soliton regime, where the 1D NLS approximation will be working again.
%We hope that the detailed information about $q(x,\t, \e)$ at the onset of the fully 3D regime will be helpful
%in tracing the solution into the soliton (genus two) regime.

The author thanks V. Kokoouline for stimulating discussions and B. Relethford, who participated in the   summer REU DMS 0649159, for   Figure \ref{Benfig}.

\end{document}